\documentclass[11pt,aps,nofootinbib,%twocolumn,
floatfix,superscriptaddress]{revtex4}
\usepackage{graphicx}
\usepackage{epsf,amsmath,amsfonts,amssymb,amsbsy}
\usepackage[mathscr]{eucal}
\usepackage{hyperref}
\begin{document}

\title{Extra dimensions and its segregation during evolution}

\begin{abstract}
{
A special isotropic energy-momentum tensor causes an anisotropic dynamical segregation of contracting extra dimensions in contrast to  expanding dimensions, provided initial data put different arrows of time evolution. Features of such the configuration are problematic and in question.
}
\end{abstract}

\author{V.V.Kiselev}
\email{Valery.Kiselev@ihep.ru; kiselev.vv@mipt.ru}

\affiliation{Landau Phystech-School, 
Moscow Institute of Physics and Technology (National Research University),
Russia, 141701, Moscow Region, Dolgoprudny, Institutsky 9}
\affiliation{State
Research Center of the Russian Federation  ``Institute for High Energy
Physics'' of National Research Centre  ``Kurchatov Institute'',
Russia, 142281, Moscow Region, Protvino, Nauki 1}

\maketitle

\section{Introduction}
%:
Trying to formulate a consistent quantum gravity reveals a problem of extra spatial dimensions  \cite{Green:1987sp,Green:1987mn} since we do not see those dimensions in any experimental phenomena. If so, we have to point to rigorous dynamical reasons or conditions for such the segregation of extra dimensions with respect to ordinary three dimensional space (3D). The most known approaches in this way are the following: i) extra dimensions are compact in contrast to the ordinary infinite space, ii) extra dimensions are large but almost empty, since the matter propagates only in three separate dimensions \cite{Rubakov:2001kp}. Both statements imply the situations when properties of extra dimensions specifically differ from that of 3D-space. In this paper we consider a mechanism providing such the difference between extra and 3D spaces. Namely, we show that initially isotropic homogeneous sources of evolution in general relativity with extra dimensions can result in an anisotropic solution when some spatial dimensions become exponentially long in comparison with others\footnote{In the similiar aspect, sources of gravity with non-equivalent dimensions were considered in \cite{Chodos:1979vk}, solutions with a cosmological constant and non-equivalent initial data were investigated in \cite{Appelquist:1983vs}, a non-liner sigma model was analysed in \cite{Ho:2010vv}, a generic case of expanding and shrifting dimensions was discussed recently in \cite{Benisty:2018gzx}.} 
if one explicitly sets different directions for the arrow of time evolution in the initial data for the velocities of expansion, i.e. the opposite signs of scale factors derivatives with respect to time in the initail data of evolution. 
%We also speculate about a mechanism providing a transition to the solution with an optimal number of such the exponentially long dimensions.

\section{Evolution}

In the isotropic homogeneous case the Einstein tensor, composed of Ricci tensor $R_{\mu\nu}$, metrics $g_{\mu\nu}$ and scalar curvature $R=g^{\mu\nu}R_{\mu\nu}$,
$$
		\mathcal{G}^\mu_\nu=R^\mu_\nu-\frac12\,\delta^\mu_\nu\,R,
$$
should have the spatial components in the form of 
$$
	\mathcal{G}^\alpha_\beta=-p\,\delta^\alpha_\beta, \qquad\alpha,\,\beta\neq 0,
$$
with $\alpha,\,\beta$ running from $1$ to $d$ being the dimension of space and $p$ denoting the pressure of matter with the energy-momentum tensor $T_{\mu\nu}$ in the Einstein equations
\begin{equation}
	R^\mu_\nu-\frac12\,\delta^\mu_\nu\,R=8\pi G\,T^\mu_\nu,
\end{equation}
where $G$ is the multidimensional analogue of Newton constant.

With the stationary tensor of matter background we expect that the pressure remains constant in time, that could be easily reached if the metrics has the form of
\begin{equation}\label{g}
	\mathrm{d}s^2= \mathrm{d}t^2-a_+^2(t)\,\mathrm{d}\boldsymbol r_+^2-
	a_-^2(t)\,\mathrm{d}\boldsymbol r_-^2,
\end{equation}
where the scale factors have the exponential dependence on the time
\begin{equation}
	a_\pm(t)=\mathrm{e}^{\pm H_\pm t},
\end{equation}
while the spatial dimensions of ``$+$ and ``$-$" components are equal to $d_+$ and $d_-$, respectively, at $d=d_++d_-$. Thus, we study the possibility of evolution with the isotropic and homogeneous initial energy-momentum tensor, while the scale factors are explicitly chosen in the form permitting the break of isotropy by setting the different initial velocities for two kinds of scale factors. Such the difference in the initial impacts to the scales would transfer the anisotropy during the evolution to produce a discrimination of spatial dimensions with essentially different properties dynamically. Indeed, the spatial components of Einstein tensor with metrics (\ref{g}) are equal to
\begin{equation}
	\mathcal{G}^+_+=-d_+H_+^2+d_-H_-H_+-\frac12\, R, \qquad
	\mathcal{G}^-_-=-d_-H_-^2+d_+H_-H_+-\frac12\, R,
\end{equation}
where the scalar curvature equals
\begin{equation}
	R=-(H_-d_--H_+d_+)^2-H_-^2d_--H_+^2d_+,
\end{equation}
while
$$
	\mathcal{G}^+_-=\mathcal{G}^-_+\equiv 0.
$$
The temporal component is given by the formula
\begin{equation}
	\mathcal{G}^0_0= -d_+H_+^2-d_-H_-^2-\frac12\, R.
\end{equation}
The isotropic condition $\mathcal{G}^+_+=\mathcal{G}^-_-$ requires
$$
	d_+H_+^2+(d_+-d_-)H_+H_--d_-H_-^2=0,
$$
that has 2 solutions: the first is $H_+=-H_-$ representing de Sitter space-time being the isotropic solution with $T^\mu_\nu=\rho_\Lambda\delta^\mu_\nu$ of vacuum with the energy density $\rho_\Lambda=d(d-1) H_+^2/(16\pi G)$ setting the curvature $R=-d(d+1)H_+^2$, while the second solution gives $d_+H_+=d_-H_-$, that is of our interest\footnote{The same relation between the dimensions and Hubble rates was obtained in \cite{Ho:2010vv} under the condition of constant physical volume at $d_+=3$ (see Section III.)}. 

Let us emphasise that in the case of de Sitter solution all of spatial dimensions do expand ($H_+>0$) or contract ($H_+<0$), and the discrimination between the regimes of expansion or contraction is set by the explicit choice of $H_+$ sign in the initial data. Therefore, we can ascribe this choice of sign to the isotropic arrow of time evolution, say, the forward arrow corresponds to the expansion or the backward arrow does to the contraction. 

In this manner, the anisotropic choice for the arrow of time evolution for the ordinary and extra spatial dimensions in the initial data could mean the anisotropic evolution of those dimensions. Indeed, the anisotropic evolution takes place at
\begin{equation}\label{eq7}
	\mathcal{G}^0_0= - \mathcal{G}^+_+ =-\mathcal{G}^-_- =\frac12\,R=- \frac12\,d\,H_+H_-.
\end{equation}
Note, that $H_+$ and $H_-$ have the same sign. Therefore, the solution in (\ref{eq7}) corresponds to the stiff matter with $\rho=p<0$. 

\subsection{Two `no-go' theorems}
The stiff matter with negative energy density could take place for a free isotropic scalar ghost field with the negative kinetic energy,
\begin{equation}
	\mathcal{L}=-\frac12\,\dot\phi^2,\qquad\dot\phi=\partial_0\phi.
\end{equation}
That is unphysical case, and we arrive to the first `no-go' theorem for the realisation of conditions in the mechanism described.

The second option is an effective field theory of scalar $\phi (t)$ depending exceptionally on the time if one has the lagrangian
\begin{equation}
	\mathcal{L}=\frac12\,\left\{1-F(\phi)\right\}\dot\phi^2,\qquad\dot\phi=\partial_0\phi,
\end{equation}
where $F$ is a function of kinetic self-interaction, that has the limit of large fields $F(\infty)=2$ and the free field limit of $F(0)=0$, say, 
$$
	F(\phi)=2\,\frac{\phi^2}{M^2+\phi^2}=2\sum\limits_{n=0}^\infty 		
	(-1)^n\left(\frac{\phi^2}{M^2}\right)^{n+1}.
$$
At $\phi^2\gg M^2$ we arrive to the stiff matter with the negative density of energy in the strong field limit. 

Then, we would get the following pattern: in the case of a restricted potential energy at large fields (a plateau of potential) the kinetic term can dominate and give the stiff matter providing  the anisotropic evolution  of space-time, but after the field reaches values with the comparable kinetic and potential energy density the regime of evolution switches to the inflation, i.e. a slow roll of field from the flat plateau to the minimum of potential. 

However, such the functions of self-interaction, $F$ changes the sign of the kinetic term, that points to the strong field limit with losing any control for the field dynamics. So, it reveals the second `no-go' theorem for the realisation of conditions in the mechanism described. 

\subsection{Option of fine tuning}
Nevertheless, one could exhibit a non-trivial situation with a composition of two fine tuned terms: 
\begin{itemize}
  \item the first term is cosmological, $T_\mu^\nu (\Lambda)=\rho_\Lambda\,\delta_\mu^\nu$ with the energy density $\rho_\Lambda>0$ and the pressure $p_\Lambda=-\rho_\Lambda$,
  \item the second term is composed by a condensate with a tuned negative energy density $\rho_{c}<0$ and zero pressure $p_{c}=0$.
\end{itemize}
\noindent
So, the condition $\rho=p$ results in 
\begin{equation}
\label{very-1}
	\rho_\Lambda+\rho_{c}=-\rho_\Lambda,
\end{equation}
hence,
\begin{equation}
\label{very-2}
	2\rho_\Lambda=-\rho_{c}>0.
\end{equation}
The fine tuning in (\ref{very-2}) points to a kind of symmetry, which should follow from some dynamical reasons unknown. In this respect, a nature of such the condensate and a mechanism of its tuning remain open, while the option itself should be pointed certainly. 

%Thus, relation of  (\ref{very-2}) composes the `very-go' conjecture for the mechanism of pre-inflationary partition of extra dimensions.

\section{Effects}
The anisotropic solution for the evolution with the isotropic energy-momentum tensor reveals two sub-spaces: the large expanding space of $d_+$ dimension and small contracting space of $d_-$ dimension. A co-moving volume $V_0$ corresponds to the physical volume
$$
	V(t)=a_+^{d_+}(t)\,a_-^{d_-}(t)\,V_0 = V_0\,\mathrm{e}^{(d_+H_+-d_-H_-)t} =V_0,
$$
 hence, the physical volume remains constant but its ``$+$'' and ``$-$'' dimensions do scale in different and opposite rates compensating each other as it was caused by the opposite choice for the arrows of time evolution in the initial data for two kinds of dimensions.

Let us consider properties of external matter evolution at fixed metrics. So, we neglect any back reaction of external matter to the metrics, that means a contribution of matter to the total energy balance to be essentially suppressed. For the sake of simplicity we focus on massless particles, i.e. on the radiation.

\subsection{Soft and hard modes}
The mass shell condition of radiation with comoving momentum $k$, $g^{\mu\nu}k_\mu k_\nu=0$ transforms to
\begin{equation}
	k_0^2=\boldsymbol{k}_+^2 \mathrm{e}^{-2H_+ t}+\boldsymbol{k}_-^2 \mathrm{e}^{2H_- t}.
\end{equation}
Modes moving exclusively along the ``$+$'' direction, i.e. at $\boldsymbol{k}_-\equiv 0$, are exponentially ultra-soft, 
$$
	k_0^2=\boldsymbol{k}_+^2 \mathrm{e}^{-2H_+ t}\to 0,
$$
while modes propagating in both the contracting ``$-$'' sub-space and expanding ``$+$'' sub-space, are exponentially ultra-hard,
$$
	k_0^2-\boldsymbol{k}_+^2 \mathrm{e}^{-2H_+ t}=\boldsymbol{k}_-^2 \mathrm{e}^{2H_- t}\to \infty.
$$
Therefore, an observer living in the large dimensions, considers the ultra-hard modes as super-heavy alike super-Planckian massive particles with masses rising during such the evolution.

We see that the energy of soft modes scales as
\begin{equation}
	k_0^+\sim\frac{1}{a_+}\sim\mathrm{e}^{-H_+t},
\end{equation}
while the energy of hard modes scales as
\begin{equation}
	k_0^-\sim\frac{1}{a_-}\sim\mathrm{e}^{H_-t}.
\end{equation}
At the constant physical volume $V(t)$ we expect that for the energy densities of soft and hard modes we get 
\begin{equation}\label{dot}
	\rho_\pm\sim\frac{1}{a_\pm}\sim \mathrm{e}^{\mp H_\pm t},
\end{equation}
that is valid. Indeed, the energy-momentum conservation $\nabla_\mu T^\mu_\nu=0$ for the temporal component, i.e. at $\nu=0$, gives 
\begin{equation}
	\dot \rho_++\dot\rho_-+(d_+H_+-d_-H_-) (\rho_++\rho_-) + d_+H_+ p_+ -d_-H_-p_-=0,
\end{equation}
where $p_\pm$ denote pressures of soft and hard modes. Since the sub-spaces are separately isotropic, but the overall isotropy is explicitly broken, we put the relations for radiation in the sub-spaces
$$
	p_\pm=\frac{1}{d_\pm}\,\rho_\pm,
$$
that results in 
$$
	(\dot \rho_+ +H_+ \rho_+ )+(\dot \rho_- -H_- \rho_-)=0,
$$
which means that the evolution of soft and hard modes can be separated and it gives relation in (\ref{dot}). 

Then, free isotropic soft and hard modes of radiation compose two cold and hot matter components with different temperatures and pressures. However, interactions could mix these components. 

Collisions of soft modes with soft modes cannot produce hard modes. i.e. the extra contracting space is decoupled from the soft $d_+$ dimensional space. Collisions of hard modes with hard and soft modes can produce both hard and soft modes, that can cause the cooling of contracted sub-space and warming of expanding sub-space. The balance depends on the rates of evolution and scattering, that determines the scenario of future. For instance, the high density of hard modes can change a geometry of extra sub-space, say, causing its compactification before the cooling, that means further complete decoupling of extra dimensions.

\section{Conclusion}
We have described the novel anisotropic solution with extra dimensions under the conditions of isotropy and homogeneity for the energy-momentum tensor with the explicit breaking the isotropy by setting different arrows of time evolution in the initial scale factors for ordinary and extra dimensions. This anisotropy discriminates expanding spatial dimensions and extra contracted dimensions.
Modes propagating in sub-spaces are differentiated in the following way: soft modes live exclusively in the expanding sub-space, while other modes look like ultra-hard and heavy for observers in the expanding world. 

Such the effects are provided by the specific option for the energy-momentum tensor that can be reached if one introduces the fine tuning for the cosmological term and pressureless condensate with the negative energy density.

A knowledge on a further fate of such the evolution itself requires to take into account an influence of additional matter to the metrics. 
Moreover, the regime described needs studies of model with deviations of energy-momentum tensor from the values constant in time. 

The segregated evolution allows us to conjecture that extra dimensions do decouple from the expanding sub-space. However, the problem of justification for the special state to produce the segregation remains open.

\acknowledgments
This work is partially supported by Russian Ministry of Science and Education, project \#~3.9911.2017/BasePart.  

\bibliography{bib_ex-seg}
\end{document}